%% file: agentplatform.tex
\documentclass[sigconf,screen,nonacm]{acmart} 

\usepackage{times}
\usepackage{url}
\usepackage{listings, lstautogobble}
\usepackage{xspace}

\usepackage[nameinlink]{cleveref}
\usepackage{hyperref}

\newcommand{\bosque}{\textsc{Bosque}\xspace}
\newcommand{\bapi}{\textsc{BAPI}\xspace}
\newcommand{\mint}{\textsc{Mint}\xspace}

\newcommand{\sundew}{\textsc{Sundew}\xspace}

\newcommand{\aisefullcaps}{Agentic Infused Software Ecosystem\xspace}
\newcommand{\aisefull}{agentic infused software ecosystem\xspace}
\newcommand{\aise}{AISE\xspace}

\newcommand{\eg}{\hbox{\emph{e.g.}}\xspace}
\newcommand{\ie}{\hbox{\emph{i.e.}}\xspace}
\newcommand{\etc}{\hbox{\emph{etc.}}\xspace}
\newcommand{\vs}{\hbox{\emph{vs.}}\xspace}

\newcommand{\cf}[1]{\texttt{#1}}

\newcommand{\explicit}{e}
\newcommand{\discover}{d}
\newcommand{\mechanize}{m}
\newcommand{\cooperate}{c}
\newcommand{\resiliant}{r}

\newcommand{\tags}[1]{}

\definecolor{cgreen}{rgb}{0.25,0.5,0.35} 
\definecolor{stringred}{rgb}{0.6,0,0} 

\lstdefinelanguage{bosque}{
keywords={},
keywordstyle=\color{blue}\bfseries,
identifierstyle=\color{black},
alsoother={@},
sensitive=true,
comment=[l]{\%\%},
commentstyle=\bfseries,
}
 
\lstdefinelanguage{smtlib}{
keywords={},
keywordstyle=\color{blue}\bfseries,
identifierstyle=\color{black},
alsoother={@},
sensitive=true,
comment=[l]{;;},
}

\setcopyright{acmlicensed}
\copyrightyear{2025}
\acmYear{2025}
\acmDOI{XXXXXXX.XXXXXXX}

\begin{document}

\title{Toward an Agentic Infused Software Ecosystem}

\author{Mark Marron}
\email{mark.marron@uky.edu}
\affiliation{%
  \institution{University of Kentucky}
  \city{Lexington}
  \state{Kentucky}
  \country{USA}
}

\begin{abstract}
Fully leveraging the capabilities of AI agents in software development requires a rethinking of the software ecosystem itself. To this end, this paper outlines the creation of 
an \emph{\aisefullcaps} (\aise), that rests on three pillars. The first, of course, is the AI agents themselves, which in the past $5$ years have moved from 
simple code completion and toward sophisticated independent development tasks, a trend which will only continue. The second pillar is the programming 
language and APIs (or tools) that these agents use to accomplish tasks, and increasingly, serve as the communication substrate that humans and AI agents interact and collaborate 
through. The final pillar is the runtime environment and ecosystem that agents operate within, and which provide the capabilities that programmatic agents use to interface with 
(and effect actions in) the external world. To realize the vision of \aise, all three pillars must be advanced in a holistic manner, and critically, in a manner that is synergistic 
for AI agents as they exist today\footnote{For the near term we expect these to remain primarily \emph{Large Language Model} LLM Transformer based}, those that will exist in the 
future, and for the human developers that work alongside them.
\end{abstract}

\maketitle

\section{Framing the Problem}

\label{sec:concepts}
This paper takes the view that creating a fully realized \aisefull requires a holistic approach to the entire software stack. Agents provide 
a powerful mechanism for understanding user intents and suggesting possible actions that, are likely, good candidates for accomplishing a task. 
However, these outputs are not strongly grounded and do not directly provide any guarantees about correctness. Without 
support from the ecosystem for discovering and managing tools, and an environment that is designed to safely execute these actions, the practical usability 
of these systems is highly constrained. To address these challenges, this work outlines a complete software stack, from the core programming 
language, to the tooling ecosystem, to the actual runtime environment, that is focused to supporting the development and operation of agentic 
software systems. 

By co-designing these components with the needs of AI agents in mind, we can create a software ecosystem that is uniquely well suited to the 
challenges of agentic software development, and that can significantly enhance the capabilities of AI agents in this space.
Thus, the first action is to outline the core design principals that concepts that guide the decisions and development of the various components of
this \aisefull. 

\subsection*{Explicit\tags{\explicit} Intents and Behaviors}
A key challenge for current AI LLM based agents is their ability to manage context and to (statistically) guess about possibly relevant information that is not explicitly 
provided in the context window. In many cases agents are able to successfully infer this information, but there is a long tail of situations where incorrect assumptions can be 
made. This compounds with the fact that many programming languages and software ecosystems have a wide range of implicit behaviors and special case semantics which creates a 
level of \emph{incidental complexity}~\cite{silverbullet,bosque} in the system. Working on code with these implicit or unexpected behaviors now involves understanding the core intent of the 
code, and then additionally, explicitly or mentally reviewing a checklist of what implicit behaviors and what special case scenarios may be present. 
This is a significant cognitive load for human developers, particularly when reviewing AI generated code, and a significant source of errors for AI agents as they must 
be aware of the full context of a codebase not just the immediate code they are working on.

With this in mind, a core design concept for \aise is to create an ecosystem where the intent and impact of code is explicitly and concisely encoded in the textual (syntactic) representation 
of the code \emph{and} that, whenever possible, corner case semantics are eliminated. As we will see, by careful language design choices, we can in fact simultaneously satisfy the 
somewhat contradictory goals of conciseness and explicitness, and in doing so, produce a software ecosystem that is uniquely well suited to the challenges faced when managing 
context window sizes and attention degradation. At the same time the pursuit of this design concept leads to a software ecosystem that is resistant to a wide range of error patterns and presents 
lower friction for human-AI cooperation.

\subsection*{Discoverability\tags{\discover}}
Context window management and tool discovery are major challenges for current AI agentic systems. There is a tension between explicitly loading anything \& everything that may be relevant 
to a task into the context window and how aggressively to prune via document retrieval and ranking heuristics. As the number of tools and context that an agent has access 
to grows, it becomes increasingly difficult for the agent to effectively manage its context and can lead to major drops in agentic performance. This problem is not unique to transformer 
based AI agents, human developers also face challenges in discovering (remembering) and managing the tools and information they need to accomplish a task! 

From this perspective it is clear that a core design concept for \aise is to create an ecosystem that simplifies the discovery of tools and information for both human and AI agents. In 
addition to merely improving discoverability, we also want the ecosystem to simplify summarization tasks and explicitly support progressive disclosure.
This is closely related to the previous design concept of explicit intents and impacts -- by making the intent and impact of code explicit, it becomes easier 
for both human and AI agents to discover which tools and information are relevant for a given task. 

Consider an API signature \cf{wait(duration: Int)} \vs the 
same API in a language that allows unit typed~\cite{fsharpunits,bsqon} primitive values, \cf{wait(duration: MilliSeconds)}. The token costs of the two signatures are nearly identical, but the second 
signature provides critical context explicitly in the signature itself while the first requires either the initial inclusion of the documentation in the context window, the agent to 
make a probabilistic guess for the unit of time, or for the agent to take the additional step to read the documentation to discover this critical information. 

\subsection*{Mechanize\tags{\mechanize} Everything}
The vision of a ``Dark Factory'' for software agents is a north-star for full automated Agentic coding~\footnote{The idea of automatic application generation is a perennial one in 
computer science~\cite{flashfill,synth1,synth2} but recent excitement in the Agentic space highlights the renewed potential in this space with LLM driven systems --
\url{https://www.danshapiro.com/blog/2026/01/the-five-levels-from-spicy-autocomplete-to-the-software-factory/} and \url{https://simonwillison.net/2026/Feb/7/software-factory/}}.
In this vision, the software ecosystem is mechanized to the point where agents can independently create entire applications without human intervention, operating merely from high level 
specifications of some form.  

Recent SotA results in this space have either used an existing application as an oracle (\ie cloning an existing application) or a full formal specification of 
the application (\ie in Lean~\cite{lean} or Dafny~\cite{dafny}). However, the assumption of an existing application or a full formal specification\footnote{In practice these 
formal specification are often as large (or larger) and as complex~\cite{sellingformal} as the code itself!} presents serious roadblocks to the realization of this vision at a large 
scale. This highlights the need to provide expressive, multi-modal and cooperative, mechanisms for specifying user/developer intent as a core concept in the \aise thinking.
By integrating specification and requirements gathering as first-class parts of the system we can significantly reduce the difficulty of expressing 
\emph{what we want to build}. By co-designing the language with a tooling ecosystem we ensure that we can then mechanically and/or cooperatively validate that the Agentic system has 
correctly \emph{built what was requested}. 

\subsection*{First-Class Cooperation\tags{\cooperate}}
The problem of specification and requirements is not just a problem for mechanization, but also a major barrier to effective cooperation between human and AI agents. The current SotA 
in human-AI cooperation in software development is to have a human provide a, mostly normative, specification of what they want to build, then an AI agent attempt to build it, and 
then a combination of testing and manual review to determine if the agent succeeded. This process is not only inefficient but also error prone, as it relies heavily on the 
human's ability to 1) write a comprehensive test suite and 2) perform careful code reviews, both of which are notoriously monotonous, difficult, and error prone tasks for human developers!.

Thus a core design concept is to integrate specification and requirements deeply into the language and system. This involves 
building effective multi-modal specification features into the programming language -- for both formal and informal specifications -- then building tooling to mechanize the 
validation of these specifications and creating workflows that provide simple/digestible feedback for a human developer to interact with.

\subsection*{Failure Safety \& Resilience\tags{\resiliant}}
Failure is an inevitable part of software development, this is true for software written by humans, and even more pressing in the context of Agentic 
software development, where the complexity and unpredictability of the system can lead to a wide range of failure modes. A core design concept for \aise 
is to create an ecosystem with multi-layered safety and resilience mechanisms. 

At the language level we want to eliminate common sources of errors and bugs, and to make it easier for developers 
to write correct code. At the system level we want to build in 
mechanisms for sandboxing resources, monitoring for data exfiltration, and managing fault logging and diagnostics. 
These mechanisims ensure that certain class of failures are impossible, or if they do occur, are constrained to safe aborts.
Beyond strict logical failures, we also consider qualitative behavior and recovery -- \eg ensuring progress even in the case of 
failure, identifying resource leaks in workflows, or workflow failures resulting in inconsistent or un-revertable states. 

\subsection*{Contributions}
\begin{enumerate}
    \item We introduce the concept of a \aisefull and outline the core design concepts that guide the development of this ecosystem.
    \item We extend \bosque with features for explicit Agent \& API interfaces as well as support for multi-modal intent specification (\Cref{sec:bsqagentic}).
    \item We describe a mechanized validation tool, \sundew, that can be used to validate the correctness of AI generated code against specifications 
    and requirements or used online by an Agent to provide formal introspection (\Cref{sec:sundew}).
    \item We outline a runtime environment, \mint, a HATEOAS~\cite{restphd} that provides progressive discovery as well as support for safety and 
    resilience when executing agentic software systems (\Cref{sec:mint}).
\end{enumerate}

\section{A Programming Language for \aise}
\label{sec:bosque}
\input{bosque.tex}

\section{Agentic \bosque}
\label{sec:bsqagentic}
\input{agenticbsq.tex}

\section{Mechanized Understanding and Validation in \aise}
\label{sec:sundew}
\input{sundew.tex}

\section{A Runtime Environment for \aise}
\label{sec:mint}
\input{mint.tex}

\section{Discussion}
\label{sec:discussion}
\input{discussion.tex}

\section{Related Work}

\paragraph{Agentic Programming Languages:} 
Despite the major role that programming language design plays in the capabilities of AI agents, there has been relatively little investigation into 
programming language design specifically for agentic software development. The most substantial work in this space 
is by Marron~\cite{bosque,bsqon} and Meijer~\cite{meijerfunctions,meijerprogai,meijeragents} who have both proposed  
designs focused on the issue of how programming language design impact the capabilities of AI programming and Agentic systems.
To the best of our knowledge the Universalis language (Meijer) has not been made publicly available or been formally described. 
The \bosque language, as well as the extensions in this work (see data availability statement), are fully open-source~\cite{bosquesrc} 
and all development is publicly accessible. The extensions in this work go far beyond the state of the art in either system, 
bringing the design from a core PL or tool, into a full ecosystem including deployment and orchestration.

\paragraph{Verifiable AI Programming:}
The topic of verifiable AI programming has been a large focus of research in the formal methods community, with significant work on using 
languages like Lean~\cite{lean} and Dafny~\cite{dafny} to formally specify and verify the correctness AI generated code. 
These systems have shown promise in verifying the correctness of AI generated code, but they also demonstrate the limits of existing 
tooling, particularly in terms of the complexity and size of the specifications required. In practice the current state of the art for 
verified AI program synthesis is limited to small, self contained programs leet-code style problems~\cite{av1,av2,av3,av4,av5}, and even then, the specifications are 
often as large (or larger) and as complex~\cite{sellingformal} as the code itself. These challenges highlight the need for simplified 
(partial correctness focused) specification languages, multi-modal support, and scalable mechanized validation tools as core components of the \aise vision.

\paragraph{Specification and Cooperation:}
The problem of specification and requirements gathering is a major challenge for effective cooperation between human and AI agents in software development. 
Of particular interest in this space is prior work on multi-modal specification and interaction with \emph{End-User Programming}~\cite{agentpreicent,flashfill} systems, such as FlashFill~\cite{flashfill} 
and Nlyze~\cite{nlyze}, and prorogued programming~\cite{prorogued}, which have shown promise in allowing users to provide specifications in a variety of forms. Related work by Gulwani et.al has 
investigated the use of examples and demonstrations as a form of specification for AI agents~\cite{uist}, which is a promising direction for reducing the difficulty of 
expressing intents (specifications) and for iterating on with AI agents in a cooperative manner. The \aise vision builds on this prior work by integrating multi-modal 
specification and validation deeply into the language and providing a platform for new UX and interaction paradigms for human-AI cooperation in software development.

\paragraph{Agentic Systems:}
Recent work on long-horizon agentic systems and tool use, such as AppWorld~\cite{appworld}, LOOP~\cite{thatrlone} and ToolFormer~\cite{toolformer}, have demonstrated the potential 
of AI agents to perform complex tasks over extended periods of time. However, these systems also highlight the challenges of context management and tool discovery, as well as the 
need for robust failure handling and recovery mechanisms as, with more complex tasks, the unaugmented success rates are at $70\%$ for simpler tasks but dropping off 
rapidly to the $45\%$ range for more complex (longer horizon) tasks that involve more complex API (tool) use. This work provides a complimentary set of contributions -- 
in the form of language design, mechanized validation, and runtime features -- that are designed to work in conjunction with improvements in raw agentic capabilities to 
create a more robust and effective ecosystem for agentic software development.

\section{Onward!}
This paper outlined the vision and core design concepts for an \aisefull, and describes the key components of this system, from the programming language to the 
mechanized validation tool to the runtime environment. Our approach to the problem of creating effective and trustworthy agentic software systems is to take a holistic approach 
to the entire software stack, co-designing the language, tools, and runtime environment to work together in a synergistic manner. As shown in this work, this approach is 
critical to addressing the challenges of context management, api/tool discovery, specification and requirements gathering, code/action generation, and safety that are critical to the success 
of agentic software systems. By advancing the state of the art in these areas, and addressing key limitations to current agentic deployment, we can create a software 
ecosystem that significantly increases the feasibility, and drives widespread adoption, of agentic software systems.

\section*{Data Availability}
Experimental versions of all of the systems described in this paper are publicly available via the main \bosque github repository \url{https://github.com/BosqueLanguage}. 
These systems are actively being integrated and converted into a production ready system that brings the \aisefull to developers (and users) everywhere.

\bibliographystyle{ACM-Reference-Format}
\bibliography{bibfile}


\end{document}

%% file: bosque.tex

We begin with a review of the \bosque language as introduced in~\cite{bosque}. As noted by the authors -- \bosque is not based on a single big feature, 
or even a number of small novel features, instead the value comes from a holistic process of simplification and feature selection with a single focus 
toward what will simplify reasoning about code -- for humans, AI agents, and formal systems. 

At the core of \bosque is a let-based functional language that is focused on eliminating the complexities associated with mutablility, aliasing, 
inductive-invariants, and nondeterminism. 
A simple \bosque program, \Cref{fig:sign}, provides a flavor of the language. The code implements a simple \emph{sign} 
function. This code is very similar to the implementation one would expect in Java or TypeScript -- in fact just 
eliminating the explicit \texttt{`i'} specifier on the literals would make it valid TypeScript. 

\begin{figure}[ht]
\begin{lstlisting}[language=bosque]
function sign(x: Int): Int {
  var y = 1i;
  if (x < 0i) {
    y = -1i;
  }
  return y;
}
\end{lstlisting}
\caption{Example \cf{sign} function in \bosque.}
\label{fig:sign}
\end{figure}

This function highlights the use of multiple updates to the same variable and block structured conditional flows. \bosque distinguishes between variables, let, that 
are fixed and those, var, that can be updated. In many languages signed numbers have asymetric ranges and thus negation is unsafe in special cases 
\ie \cf{-INT\_MIN} will error or silently wrap. However, in \bosque the dynamic ranges for signed integers are symmetric, and aligned with their unsigned versions as well, 
so that negation is always safe and corner case issues with signed/unsigned conversions are eliminated\tags{\explicit,\resiliant}.

Another distinctive feature of \bosque is the complete elimination of loops for container processing. Instead, \bosque provides a rich set of higher-order functions for operating 
on collections -- similar to Java Streams or C\# LINQ as shown in \Cref{fig:listbosque}.

\begin{figure}[ht]
\begin{lstlisting}[language=bosque]
let l = List<Int>{1i, 2i, 3i};

l.allOf(pred(x) => x >= 0i) <@\%\%@> true 
l.map<Int>(fn(x) => x + 1i) <@\%\%@> List<Int>{2i, 3i, 4i}
\end{lstlisting}
\caption{Eliminating the need for loops using higher-order functor operations in \bosque.}
\label{fig:listbosque}
\end{figure}

The use of higher-order functions allows for a natural, concise, and explicit way to express collection processing\tags{\explicit,\discover}. Most directly the representation is strictly more token efficient than the 
equivalent loop-based code \emph{and} we avoid the need for the agent to repeatedly generate common loop indexing and control flow, with the associated risks of off-by-one errors, incorrect variable name 
selection, and inverted conditions, that are regular occurrences in these types of code~\cite{sstubs4J}\tags{\resiliant}. Thus, the use of higher-order functions (and \bosque) not 
only supports the goal of token efficiency, but also improves the probability of successful (correct) code generation!

This approach also allows for a more direct and explicit expression of intent. Again, from the perspective of agentic code generation, the mapping from latent intent to semantically meaningful
operation names improves the likelihood of of correct operation selection and, when ``reading code'' enables the agent to simply focus attention on a single operation name instead 
of needed to analyze all components in a multiline looping implementation. Interestingly, human developers also benefit from this increased clarity when reading and reviewing code\tags{\cooperate}, 
as, they can also immediately understand the intent of the code without parsing through control flow details. For example, the use of \texttt{allOf} makes it clear 
that the intent is to check if all elements satisfy a condition, which is more explicit than a loop with a conditional check and a break statement.

The final feature of \bosque that we highlight here is the ability to easily create strong type aliases and enforce invariants on, both aliased and composite, data values. Consider 
the code in \Cref{fig:alias-invariants} that declares two type aliases, \cf{Fahrenheit} and \cf{ZipCode}, and two composite entities, \cf{TempRange} and \cf{TempForecast}.

\begin{figure}[t]
\centering
\begin{lstlisting}[language=bosque]
type Fahrenheit = Int;
type ZipCode = CString of /[0-9]{5}('-'[0-9]{4})/c;

entity TempRange { 
  field low: Fahrenheit; 
  field high: Fahrenheit; 

  invariant $low <= $high;
}

entity TempForecast {
  field location: ZipCode;
  field temp: TempRange;
}
\end{lstlisting}
\caption{Examples of types aliases and invariants in \bosque.}
\label{fig:alias-invariants}
\end{figure}

The code in \Cref{fig:alias-invariants} shows how \bosque allows developers to create strong type aliases that provide explicit semantic identity\tags{\explicit} to otherwise primitive values. 
This allows the developer, or AI agent, to express intent more clearly. For example, using \cf{Fahrenheit} instead of \cf{Int} makes it clear 
that the value represents a temperature (in Fahrenheit)\tags{\discover} and prevents unit-confusion~\cite{fsharpunits} or argument confusion bugs~\cite{paramswap}\tags{\resiliant}. As we will see in \Cref{sec:bapi}, 
this is very useful when specifying APIs where communication formats\tags{\cooperate} heavily use primitive values. 

Beyond simple type aliasing, we can also attach explicit invariants to both aliased and composite data types. In this example, the \cf{ZipCode} alias uses a regular expression to 
declare that a \cf{ZipCode} value is a string \emph{and} to be valid it must match the specific pattern\footnote{The regular expression language used by \bosque is a specialized 
design targeted to eliminating common forms of ReDoS attacks\tags{\resiliant} on validation~\cite{redos}}. Similarly, the \cf{TempRange} entity declares an invariant that the \cf{low} 
field must be less than or equal to the \cf{high} field. These invariants can be checked at compile time or later as dynamic checks or in the \sundew validator\tags{\mechanize}.

This ability to explicitly specify intent and invariant properties as part of the type system greatly enhances both the ability to reason about code and also to catch errors early in 
traditional development. It also provides a powerful mechanism for AI agents to generate code that is more likely to be correct, as the invariants can be used to guide both 
the initial generation, by making the constraints explicit for the LLM, and also used to provide feedback with failures of test-cases or static analysis systems.

The validation of string structures is particularly important when APIs and data formats heavily use strings to represent structured data. Tracking the possible 
data content, including sensitive PII or un-sanitized user controlled input, is otherwise a difficult problem that requires reasoning over flows 
across the entire program. With oversights leading to potentially serious security vulnerabilities~\cite{mitre2024top25} like SQL injection or 
leaks of sensitive data. 

These examples highlight the core design philosophy of the \bosque language and how it supports the goals of \aise. By eliminating 
complexity and providing powerful abstractions, \bosque enables developers and AI agents to write code that is easier to generate \& reason about as well as providing multi-layered 
support for safety \& fault detection.

%% file: agenticbsq.tex

Fully integrating the \bosque language with agents and including them a first class part of the system requires additional extensions. As currently 
described, \bosque is a general purpose programming language and an excellent target for agentic code generation but lacks explicit facilities for 
calling and orchestrating sub-agent workflows (\Cref{sec:apis}) and does not provide support for explicit decomposition and modularity of development 
tasks (\Cref{sec:holes}). 

\subsection{Explicit \cf{api} and \cf{agent} Calls}
\label{sec:apis}
In an \aisefull, we expect agents and deterministic workflows to be first class citizens in the system. Thus, we need explicit language features 
for agents to call workflows, workflows to call agents, and agents to call other agents.

We introduce two new language constructs, \cf{api} and \cf{agent}, that allow for explicit calls to deterministic workflows and agents, respectively. 
The \cf{api} construct supports calls to deterministic workflows that are defined elsewhere in the system or on remote (RESTful / HATEOAS) style 
endpoints\footnote{\Cref{sec:bapi} describes the BAPI system, which provides a powerful framework using \bosque types as a literal interchange format for calls/tools and for 
progressive discovery in \Cref{sec:mint}.}. The \cf{agent} construct allows for explicit calls to agents that are defined elsewhere in the system.
In theory these constructs could be combined into a single \cf{call} construct but this design, with the distinction between them, allows us to 
provide additional language support for the more free-form textual interaction that exists with \cf{agent} calls.

\begin{figure}[t]
\centering
\begin{lstlisting}[language=bosque]
%** Transfer amt from payer account to payee account. **%
api transfer(amt: USD, payer: Account, payee: Account)
  env={
    PAYMENT_AUTHORIZATION: OAUTH_TOKEN,
    PAYMENT_LIMIT: USD
  }
  permissions={
    \account:${payer.routing}/${payer.account}\
  }

  requires 0.0<USD> < amt;
  requires amt <= env.PAYMENT_LIMIT ||
    $events.contains(Approve{|payee=payee, amt=amt|});
;

%**  
* Given a natural language (plaintext) message,
* compute the amount to pay and send to the payee.
**% 
action splitBill(msg: String, payee: Account) {
  let amt = agent Chat::compute<Option<USD>>(
    env{}, msg, "What is half of the bill?"
  );

  if(amt === none) {
    return fail("Could not get amount from message.");
  }

  api transfer(env{...}, amt, env.account, payee);
  ...
}
\end{lstlisting}
\caption{Example of an \cf{api} defintion and an action using \cf{api} and \cf{agent} calls in \bosque to perform a payment transaction -- \eg 
\cf{splitBill("lunch was \$45.50", contacts.get("Tom"))} with the expected result is a successful payment transaction of $\$22.75$ to Tom's account.} 
\label{fig:api-and-agent}
\end{figure}

\Cref{fig:api-and-agent} shows an example of these constructs and how they can be used as part of a workflow that takes in a natural language message, say from an email 
or text with a dollar amount, splits the bill in half and then performs a payment transaction. 

The first step is to use a chatbot AI agent to extract the amount from the message via the \cf{Chat::compute} operation. The actual implementation of this 
agent is entirely open, could be remote or local, and could use any number of techniques to extract the amount. The key to this design is the agent signature which allows us to 
export it into a \bosque program, and then a packaging system which links in the needed module or sets of the remote invocation\footnote{This design enables us to make 
agents modular, versionable, and we can support multiple side-by-side models and personas just as with code packages -- \ie NPM~\cite{npm}}. 

This \cf{api} (\cf{agent}) invoke statements allows concise and explicit descriptions for what an agent should be computing and what it may access.  
Fundamentally, apis and agents are always operating in some context or environment. This context may include ambient information, such as the current 
location, or in this case of \Cref{fig:api-and-agent}, a limit on spending. The api/agent design allows us to explicitly describe 
this context as part of the api/agent definition and to specify the exact values provided at invocation. In this example, the agent is explicitly 
given an empty environment, forbidding it from accessing (and potentially accidentally or maliciously exfiltrating) any information from the 
ambient context, such as a secret \cf{PAYMENT\_AUTHORIZATION} token.

Additionaly, making the \cf{agent} invocations explicit\tags{\explicit,\cooperate} enables a simple syntax for converting un-structured (or semi-structured) outputs from an AI 
agent into a structured format. In \Cref{fig:api-and-agent} the agent is asked to compute a value for ``half of the lunch bill'' from a free-form string, and as it is a \cf{Chat} agent, 
it will by default produce a free-form string output. However, the \cf{agent} call explicitly supports providing a result type shaping signature, in this case \cf{Option<USD>} (a type alias of \cf{Decimal}). 
In general this can be accomplished by running the \bapi parser directly on the output of the agent, but if supported by the agent implementation, this can also be used to 
drive structured~\cite{picard} (or even typed~\cite{vechevtyped}) output generation without requiring additional data format specifications.

Once the amount is extracted, the workflow uses the \cf{transfer} API to perform the actual payment transaction. The \cf{transfer} API is defined with explicit environment 
variables and permissions that are required to perform the transaction, as well as preconditions that must be satisfied for the transaction to be successful. The \emph{api} 
specification includes the environment variables that are required for the API to function, such as the \cf{PAYMENT\_AUTHORIZATION} token and the \cf{PAYMENT\_LIMIT}, as well 
as the resources that the API will access, in this case the payer's account information. By explicitly declaring these requirements\tags{\explicit}, we ensure that the agent 
has the necessary context and also prevent it from accessing\tags{\resiliant} any information that is not explicitly provided. 

The API specification also includes preconditions that must be satisfied for the API call to be successful. 
In our example the \cf{api} has the precondition that \cf{requires 0.0<USD> < amt}. However, in an \aisefull, we 
also need to be able to express and check temporal properties that involve the sequence of events that occur 
during the execution of a system. In our example, we want to ensure that either the payment amount does not 
exceed a certain limit \emph{or} that in some previous event we obtained explicit user approval event for the operation.

To handle this we introduce the concept of an \emph{event log} that is implicitly tracked as part of the execution of the system. This event log can be used to record any events 
that occur during the execution of the system, such as user interactions, API calls, or agent calls. These events are inserted into an otherwise immutable log by the system, 
preventing any possibility of tampering byt developers or agents, and this log can then be referenced in the pre/post conditions of API calls to check for temporal properties. 
In our example, we can check if the event log contains an explicit user approval event for the given item. The 
ability to explicitly\tags{\explicit} express properties these requirements and conditions, are a critical foundation for later mechanized 
validation\tags{\mechanize} and runtime safety\tags{\resiliant}. 

This explicit information also provides improved discovery\tags{\discover}, as critical API information is now in a structured and explicit form, which supports improved 
agent code generation and tool use, as the agent can directly reference the API specification to understand the requirements for successful use. 
Finally, this system design opens the possibility for more advanced feedback and learning systems, as the system can provide detailed 
information on why an API call failed, such as unsatisfied preconditions. This information can further be integrated with validation and reasoning tools into the agent's training 
process via direct reinforcement learning algorithms with tools~\cite{thatrlone,retool,artist} as well as providing immediate feedback to the agent during training rollout -- allowing us to 
generate action rewards without a full trajectory.

\subsection{Holes and Meta-Thunking}
\label{sec:holes}
A critical problem for agentic code generation is modularity and decomposition. In traditional software development, developers can break down complex problems into 
smaller, more manageable tasks which can then be handed off (or deferred) for later development. This is critical for managing complexity and enabling collaboration. Strangely, 
programming languages do not, in general, provide explicit language features to support this. Instead, developers have to rely on external tools 
such as issue trackers and markdown documents, or use adhoc patterns such as writing incomplete functions with \cf{TODO} comments and aborts to indicate that a particular task 
needs to be completed later. This is a major gap in the software development process, and it becomes even more problematic when we introduce agents into the mix. 

We have extended the \bosque language with an explicit \cf{hole} construct~\cite{typedholes,nlyze,sketch} that allows developers and agents to explicitly indicate that a particular task or 
piece of code is incomplete and needs to be filled in later. From a syntactic standpoint this provides a clear and explicit\tags{\explicit} way to indicate that a particular task 
is incomplete, and it also allows us to provide additional metadata about the task, such as the expected input and output types, the context in which the task should be completed, 
and any relevant information that may be helpful for completing the task. From the standpoint of pre-training, this feature also has the potential to support explicit specialization 
of models for agents, architect \vs code generation, as holes are now representable by explicit tokens in the training data.

\begin{figure}[t]
\centering
\begin{lstlisting}[language=bosque]
function sign(x: Int): Int {
  var y = 1i;
  if (x < 0i) {
    y = ?_ -> Int;
  }
  return y;
}

%** Compute the absolute value for given integer **%
function abs(x: Int): Int 
    ensures $result >= 0i;
{
  ?_absbody(examples = true);
}
\end{lstlisting}
\caption{An example of using hole expressions in the sign function to indicate that the implementation of the negative number case is incomplete. The hole expression 
can be a simple un-named expression, with a desired type, or alternatively, can include a doc-comment, a name, and even information on where to find 
desired input/output examples.} 
\label{fig:holes}
\end{figure}

\Cref {fig:holes} shows an example of using the \cf{hole} construct in a simple function that computes the sign of an integer. In this example, we have implemented the positive and 
zero cases, but the logic to complete the negative case left as an expression-hole. By using the \cf{hole} construct, 
we can explicitly indicate that this part of the code is incomplete and needs to be completed later -- without requiring additional annotations or assertion style hacks. 

The final part of \Cref{fig:holes} uses a hole for the entire function body and, using the post-condition, can 
specify the expected behavior of the implementation along with the docstring and examples for the hole to further specify the desired behavior. This is a powerful way to provide explicit 
specifications for the desired behavior of the code, and as shown in \Cref{sec:sundew}, can be used to symbolically validate the behavior of any generated implementation.

This explicit support for modularity and decomposition is critical for agentic code generation, as it allows agents to break down complex problems into smaller, more manageable tasks 
that can then be completed later, either by the same agent or by a different agent. It also provides a standard structure for embedding additional meta-information about the task which 
enables improved discovery and more effective code generation. The direct integration also enables integration into development tools and runtimes. For example, we can trivially link 
in the concept of \emph{Prorogued Programming}~\cite{prorogued} or and LLM extended version of meta-thunking.

In this design, when a hole is executed, instead of throwing an error, the system can automatically trigger a call to a human developer or agent to complete the hole. In this workflow, 
the human (or agent) can manually specify the correct result type, in our case if the value of \cf{x} is \cf{-5}, we would provide \cf{-1i}. In later executions, the system can use these 
as memoized results for the hole, and if there is an examples file, can load/store these examples for later reference -- as unit tests or guides for an agent during code generation.

%% file: sundew.tex
In contrast to other widely used languages\footnote{A notable exception is Morgan-Stanley's Morphir~\cite{morphirrepo,formalmorgan}.} where the semantics are 
not amenable to mechanized analysis, due to features like loops, mutability, and non-deterministic behaviors, or languages like Lean~\cite{lean} and 
Dafny~\cite{dafny} that provide support for full functional verification but require substantial proof engineering expertise, \bosque is designed 
to support fully-mechanized reasoning and validation.

\subsection{Mechanized Understanding and Validation}
As described by the \bosque developers~\cite{formalmorgan}, the design of \bosque enables us to map it, almost entirely, to efficiently 
decidable theories supported by a SAT-Module-Theory (SMT) solver~\cite{z3}. Operations on numbers, data-types, and functions all map to core 
decidable theories -- Integers, Bitvectors, Constructors, Uninterpreted Functions, and Interpreted Functions. In SMT solvers the theories of Strings and Sequences 
are semi-decision procedures in the unbounded case. However, we can bound the sizes of these kinds as inputs and then the system becomes fully (and efficiently) 
decidable. As a result validation can be fully automated. There is no need for developers to learn an additional proof language, all checks are 
encoded as assertions, pre/post-conditions, and invariants in the \bosque programming language, and no manual intervention is required to write lemmas 
or diagnose proof failures!

Given this design, consider the sample \cf{sign} function from \Cref{fig:sign} and the resulting SMTLib~\cite{smtlib} encoding, \Cref{fig:sign_smtlib}. 
This code is a fully decidable translation encoding of the \bosque version. 

\begin{figure}[ht]
\begin{lstlisting}[language=smtlib]
(define-fun sign ((x Int)) Int
  (let ((y 1)) 
    (ite (< x 0) (let ((y -1)) y) y)
  )
)
\end{lstlisting}
\caption{Sign function automatically converted into SMTLib.}
\label{fig:sign_smtlib}
\end{figure}

The \bosque language includes builtin support for various validation workflows using this encoding strategy, including a specialized declaration of 
a parametric property test. A simple validation test that checks that the \cf{sign} function returns a value in the 
range $[-1, 1]$ for any input (\cf{x}) is shown in \Cref{fig:check_test}. For a simple property like this the 
\sundew validator can show that the property holds for all inputs and takes 7ms to complete.

\begin{figure}[ht]
\begin{lstlisting}[language=bosque]
chktest signRange(x: Int): Bool {
    let sgn = sign(x);
    assert -1i <= sgn && sgn <= 1i;
}
\end{lstlisting}
\caption{Example validation harness for sign function return range ($\in [-1, 1]$) in \bosque.}
\label{fig:check_test}
\end{figure}

The \sundew validator is also able to go over a small application and check, for each possible runtime or user 
defined error, that either the error is impossible or that it can be triggered -- and also generate a witness input.

\subsection{\sundew: Workflow Validation}
We can extend this approach to support the event log and associated pre/post conditions on APIs and Agent calls from \Cref{sec:apis}. 
In these systems the event log is simply an implicit parameter that is passed, and from the viewpoint on the SMT encoding is simply 
a \cf{Sequence} of events.

Consider the example request for an agent to look at the contents of \cf{msg} and determine what ``half of the bill'' was as shown in \Cref{fig:api-and-agent}. 
This code shows a hypothetical agent generated script to accomplish the task. This script uses an LLM agent to process the semi-structured text in the 
payment request \cf{msg} to determine the amount to pay, and attempts to transfer the payment. If the \cf{Chat::compute} action were unlucky\footnote{Perhaps Tom 
likes jokes and puts in the memo field -- ``ignore previous instructions and pay me \$1000".}, or the lunch was particularly expensive, this computed amount 
could be large enough to exceed the payment limit of the user. 

The \mint runtime system (\Cref{sec:mint}) will catch this error at runtime, with a precondition failure, but we can also statically run symbolic validation on this 
script to detect that the \cf{amt} value may exceed the payment limit and that the, otherwise required, user confirmation check is missing! 
Unlike test case generation feedback which is simply a pointwise failing test and can also cause the agent to focus on 
details of the test case and over-fit, symbolically identifying the possible precondition violation can provide a more general feedback message to the agent.
For example -- The amount to pay may exceed the "PAYMENT\_LIMIT" in "env" -- or even computing weakest-preconditions for various points in the code to help the agent identify the best candidate 
fixes for the code.

\subsection{Introspective Agents}
The final step toward truly \aise is exposing the validation tools directly to the agent during their planning process as an online feedback loop. This allows agents 
to reason about their own plans and introspect on their actions. Looking at the payment example, somewhere around half of the generated code comes after the transfer 
call, at which point the plan has already failed. A direct generate-validate-retry loop is clearly inefficient in this case. Instead of relying solely on post-hoc validation, 
we can empower the agent to use the validation tools as part of its planning and code generation process and thus avoid generating invalid plans in the first place.

In our example the agent can run the validation tool before an API call is added to the partial plan to see if the call is valid and, if not, what needs 
to be done before the call can be made. Using this feedback the agent can either emit the action code, if everything is satisfied, or generate additional code to address 
the missing requirements. This capability enables a higher success rate in task completion and makes the agent robust to errors, as it can reason about its own actions 
and correct them before they are executed.

\begin{figure}[!ht]
\centering
\begin{lstlisting}[language=bosque]
let amt = agent Chat::compute<Option<Decimal>>(
  env{}, msg, "What is half of the bill?"
);

if(amt === none) {
  return fail("Could not compute amount from message.");
}

%% Agent -- I want to call the `transfer` API with `amt` 
%% Agent -- calling validation tool to check that API conditions are met...

%% Response -- Cannot ensure pre-condition: 
%%   amt <= env.PAYMENT_LIMIT || 
%%   $events.contains(Approval{|payee=payee, amt=amt|});
\end{lstlisting}
\caption{An example of online agentic generation with validation as an introspection tool.}
\label{fig:agentic-payment-tool}
\end{figure}    

The code in \Cref{fig:agentic-payment-tool} shows an example agent \& tool chain-of-thought, tool use, and response. At the point where the agent is preparing 
to call an API, \eg the \cf{transfer} API, it can call the validation tool to check that the requirements for the API call are met. The validation tool can then
respond with a list of missing checks or requirements that the agent needs to address before the API call can be made. Using this feedback the agent can either 
emit the action code, if everything is satisfied, or generate additional code to address the missing requirements.

This capability enables the agent to achieve a higher success rate in task completion and be more robust to errors, as it can reason about its own actions and 
correct them before they are executed. Further, this approach has the potential to enable a new class of reactive, notebook style, agents that interleave code 
generation, validation, execution, and user interaction, to accomplish complex tasks.

%% file: mint.tex

The previous sections described a language for computation, a means to express intents in formal ways, 
and a workflow for validating and understanding the behavior of a system. 
The remaining issue is how to actually deploy and operate these systems. In this section we review the 
design of \bapi, a protocol APIs and agent system interactions, and \mint a runtime ecosystem for discovery, 
deployment, and operation of workloads in an \aisefull.

Agents are a unique new class of workloads where, in some cases the platform will be running a fixed workflow or 
agentic task, in others the agent will be exploring available services and taking actions in an incremental and 
online fashion. In the second case we, our system is not just a static execution graph, but must support dynamic 
discovery, invocation, and progressive exposure of services. Interestingly, this model has two clear precedents 
in the form of COM and HATEOAS~\cite{restphd} -- both of which were designed to support dynamic 
discovery and invocation of services. Drawing from the principles and lessons of these system, this section describes 
a novel Agentic HATEOAS model for system \aise architecture. 

\subsection{\bapi -- Bosque API Protocol Review}
\label{sec:bapi}
The first component of this system is the \bapi protocol for describing APIs and agent interactions as well as a 
means to encode and transport literal values and data structures. The design of \bapi is described in detail 
in~\cite{bsqon} and we review the key features here.

As with the \bosque language, the design of \bapi is focused on supporting mechanized correctness and explicitly encoding 
intents (and key information) into the syntax of API \& data definitions. In addition, \bapi provides a literal syntax 
for encoding data that is more efficient (token-wise) than JSON, is fully round-tripable, and designed to be easily authored 
by humans (or AI agents).

\begin{figure}[t]
\centering
\begin{lstlisting}[language=bosque]
%** Definition of an Order type w/ sensitive TIN **%
type OrderId = CString of /[A-Z][0-9]+$/;
sensitive type TIN = CString of /[0-9]{9}$/;

entity Order {
    orderid: OrderId;
    amount: Decimal;
    customer: TIN;
}

%** Literal Order (Complete Form) **%
Order{
    orderid = 'A53'<OrderId>,
    amount = 45.50d,
    customer = '123456789'<TIN>
}

%** Literal Order (Token Minimized Form) **%
Order{
    'A53',
    45.50d,
    '123456789'
}

%** Literal Order (Standard Form w/ Leakage Filter) **%
Order{
    'A53'<OrderId>,
    45.50d,
    '*********'<TIN>
}

%** Literal Order (JSON) **%
Order{
    "orderId": "A53",
    "amount": 45.50,
    "customer": "123456789"
}
\end{lstlisting}
\caption{Example of an \cf{Order} type definition and $3$ literal value representation in \bapi, complete, token 
minimized, and with sensitive leakage filtering applied -- plus the automatically 
generated JSON equivalent representation.} 
\label{fig:order-example}
\end{figure}

\bapi specifications are a extension of the \bosque type system. A simple example of a \bapi specification is shown in 
\Cref{fig:order-example}. The \bapi specification defines a set of types that make up a customer \cf{Order}. 
The first declaration is a simple type alias for an \cf{OrderId} that is a structured string (matching a regular expression) 
while the second declaration is for a \cf{sensitive} type for a \cf{TIN} (Taxpayer Identification Number). The second 
declaration shows how \bapi supports information control and monitoring by explicitly marking data as \cf{sensitive} 
so that later uses and operations on the data can be mechanically checked for compliance or security risks.

These two alias are then used in the composite \cf{Order} type that also includes an \cf{amount} field. Finally, 
the example shows various \bapi options for serializing literal \cf{Order} values. All of these forms + parsers and 
serializers are automatically generated from the type definitions and the choice of which form to use is up to the 
user (or agent) based on the needs of the scenario. By design they can be intermixed in the same system and are 
amenable to linear time processing and can be stream/zero-alloc parsed.

In the first form, the literal value is fully annotated with type information. This form is ideal for mechanized 
reasoning and validation of the data but, like JSON, is verbose and can involve massive redundancies in property names 
(or other tags) which increases token load and message sizes. The second form is a token-minimized encoding that relies 
on the order of fields in the type definition to avoid repeating tags and type information. This form is ideal for 
efficient encoding and transport of data. 

The ability to use (and intermix) both forms allows users and agents to work with high-information representations for 
convenience, or when feeding (small) results of a tool-call into an agent context, while using a token-minimized form for 
transport, storage, or losslessly compacting large values for agent context management.

The third form is the same as the first form but with a leakage filter \emph{automatically applied} to the \cf{TIN} 
field which was marked sensitive. This form is the default for emitting data in scenarios like logging or crash dumps. 
This minimizes the risk of side-channel leakage and, if working in an untrusted agent, provides a means to 
guarantee sensitive information cannot be leaked. 

The final form is the JSON representation that is automatically generated from the type definition. This allows interop 
with other systems and tools even if they do not support \bapi natively. By design, the JSON form is fully round-trippable 
with the encodings that avoid common \cf{NaN}, \cf{MAX\_SAFE\_INTEGER}, date format, \etc handling issues. This allows 
our \aise stack to work seamlessly with existing tools and systems while enjoying the benefits of the more efficient \bapi 
specification and literal formats.

\subsection{Info Routes and Progressive Discovery}
\label{sec:mint-discovery}
The original formulation of REST and HATEOAS~\cite{restphd} envisioned systems where the description of the platform 
capabilities was inline with the ability to dispatch these operations. Further, that a user could discover and 
progressively explore details of these capabilities and how to invoke them.

However, modern service based architectures and platforms, drawing from the heritage of programming languages, split the 
executable artifact from the documentation and APIs. Further, our objective involves supporting multiple modes of 
interaction with a single logical service, \ie both \bapi (Verbose, Minimal, or even Binary Encoded) and JSON. In addition 
we want to provide build and deployment systems that declaratively generate all of the appropriate bindings, set the 
needed routes, setup ``well known'' names for discoverability, provide integrated search, and are optimized for 
agentic interaction. To support this we introduce a novel runtime platform, \mint, to handle the configuration and 
execution of \aisefull services.

A deployed \mint server is setup around a \bapi configuration file that specifies routes based on URI globs. Each 
route may be a static file (for simple resources) or connected to a \bosque task, and of course middleware options 
for logging, authorization, and request/response meta-data management -- very similar to popular frameworks 
like express.js~\cite{express}. The \mint server can also automatically setup support for multiple encodings (Verbose, Minimal, Binary, JSON) along with 
special routes for discovery and info. By convention \mint sets a top-level route named \cf{/actions} that returns a compact structured specification 
of all of the endpoints available on the server, filtered by the permissions of the caller, which include signatures of the handlers, pre/post conditions 
and documentation comments. Each endpoint is also setup with a \cf{/actions/\{endpoint\}} route that returns more detailed information about the endpoint 
including normative usage patterns, examples, URI links to related resources, and more detailed documentation. This allows for \mint to satisfy the primary use cases of systems like 
MCP~\cite{mcp} or Skills~\cite{skills} including autonomous \& progressive discovery of the capabilities of the service in 
a structured, secure, and efficient manner.

To further support agentic use cases, \mint also sets up a \cf{/search} route that accepts a query, as plain text, and is intended to facilitate semantic search over the service's 
capabilities and related information. By default, the search route is implemented to perform a structured semantic search over the specifications of the endpoints and documents 
but the design of \mint allows for this to be customized and to include links to other related services or resources (\eg package manager or websites). This allows agents to ask 
questions about the service without prior knowledge and autonomously obtain relevant information for a task without.

This opinionated design ensures a standardized way to discover the features and functionality a \mint service provides. 
The structured \bosque specifications that are returned ensure that an agent can mechanistically understand the usage 
of a given feature \emph{and} the ability to incrementally query for more details allows for the provision of detailed 
information, use examples, and normative patterns without overwhelming the agent context. Further, by integrating 
search as a dedicated route we can support powerful agentic workflows with a strong seperation of concerns -- specifically 
the agent is responsible for deciding what concepts are relevant to the task at hand while the \mint runtime is responsible 
determining the best way to find and return relevant information. 

\subsection{Exposure, Sandboxing, and Guard-Rails for Agentic Workloads}
\label{sec:mint-box}

Once we are running agentic workloads we need to manage the unique security and operational challenges they present. 
Again our system is able to leverage the design of \bosque and \bapi to mechanistically understand the behavior of these workloads 
and monitor them for compliance and security.

The first level of protection is via the \bapi specifications themselves. Each API specifies the types of data it accepts and returns, 
including information on sensitive data. Thus, \mint automatically checks that sensitive data is not sent over an public endpoint
(by default all endpoints are private) and private endpoints can also be flagged as no-sensitive (or specific levels). This allows for 
immediate linting of endpoints that may publicly expose sensitive data and also allows for analysis of any workflow to determine if 
it can access sensitive data, and if so, what data and what the potential risks are. For example, when an agent accesses and endpoint 
with sensitive data, the runtime can automatically quarantine any outputs from that agent which may now be tainted with sensitive details.

In addition to monitoring data-flows and sensitivity, the \mint runtime also automatically sandboxes execution based on the URI resources 
it has declared it will access. In \Cref{fig:api-and-agent} the \cf{transfer} task is declared to access the \cf{account} resource for the 
payee alone. Thus, the implementation is unable to, maliciously, access the ballance or or move money from the payer's account. This technique 
can also be applied to any resource that is mappable to a URI, for example using the glob syntax to whitelist a certain set of REST resources 
to prevent data exfiltration to an unauthorized endpoint or prohibiting access outside of the \cf{file:///tmp/app\_name/} folder to allow intermediate 
files to be written/read while ensuring that user data is not accessed. This ensures that even if an agent is compromised or behaves maliciously, 
it cannot access resources outside of its declared scope.

While the idea of sandboxing is well known, the key insight here is using URIs and Globs as the means of resource specification and sandboxing. 
As opposed to the historical use of custom resource types and access languages~\cite{javaresource,bsdjails,ztdjava} which have, historically, been 
difficult to standardize, understand, and use, URIs and globs naturally match the model of resources used in RESTful systems. This allows for 
a simple, uniform, and powerful way to specify and enforce resource access policies across a wide variety of resource types (files, REST endpoints, 
databases, \etc) without needing custom plugins or extensions for each type. 

Finally, the \mint runtime also provides a dynamic monitoring and enforcement system for agentic workloads. As \bosque and the \bapi specifications 
efficiently executable pre/post condition, invariants, and assertions, these can all be monitored and enforced at runtime by the \mint system. Failures 
of any of these conditions triggers a safe-abort of the offending task, and can be configured to trigger additional mitigation such as rolling back state 
changes, logging relevant data, and even producing a offline time-travel-debuggable dump~\cite{ttdjs}. This provides a powerful line of 
defense against misbehaving agents.
This multi-layered and explicit approach to security and safety implies that, even if a malicious actor is 
able to exploit and work around one layer of protection, then they are still limited by the additional layers 
either blocking the exploit or forcing attackers to find multiple bypasses for an exploit to chain.

%% file: discussion.tex

A key question is whether, in the future, a sufficiently powerful AGI agent will even need or benefit from 
the features of the \aisefull or if, by sufficient training and power, it will be able to operate on any language/platform?

Consider a hypothetical limit where we assume that an eventual AGI agent is as capable as an expert human developer, 
top 10\% in the world, then we can answer this question by asking the analogous question of whether this expert would be aided by the features 
outlined in this work. The answer is yes, they would, and in fact, for many of the features we have outlined part of the description and motivation 
is how it eliminates or helps with problems real developers have today.

When going for three (or more) $9$'s of reliability then even small possibilities for 
failure ($1\%$ or less) are unacceptable as, at this level, almost any mistake would exceed the failure budget. Would we accept a $1\%$ chance that 
an agent would stop paying our mortgage, delete a customer database, post medical information to a public forum? Thus, when working to build 
highly reliable and trustworthy agents we need to have multiple overlapping layers to drive reliability and it is ill-advised to forgo any possible 
advantage. 

The converse question is whether the design and principles outlined here are overfit 
for the current state of the art in LLMs and attention-based models. When these models change or are replaced with new architectures will the 
design of this system become obsolete or even a hindrance?

As with the previous question, the first response is to again consider the hypothetical limit of replacing an LLM with an expert human developer. 
Again we see that, the principles outlined here are not specific to attention-based models or LLMs, but generalize to humans, as well as earlier 
(simpler) program synthesis systems (\eg NLyze~\cite{nlyze}). 

Additionally, the system design is set to allow agents to be swapped out transparently. Features like the \cf{agent} invokes 
are explicitly designed to avoid committing to plain text as arguments and the \sundew verifier work with any code regardless of who/what produced it. 
In practice an agent could be a human in the loop and the proposed \aise would still be effective and coherent. Thus, the proposed design will remain 
viable and effective even as the underlying technology evolves.